\documentclass[12pt]{iopart}

\usepackage{iopams} 
\usepackage{cite}
\usepackage{color}
\usepackage{graphicx}
\usepackage{setspace}

\usepackage{wrapfig}
\usepackage{comment}
\usepackage{caption}
\usepackage{placeins}
\usepackage{mdframed}
\usepackage{mdwlist}

\usepackage{hhline}

\begin{document}

\title[]{Modelling of vorticity, sound and their interaction in two-dimensional superfluids }

\author{Stefan Forstner$^1$, Yauhen Sachkou$^1$, Matt Woolley$^2$, Glen I. Harris$^1$, Xin He$^1$, Warwick P. Bowen$^1$ and Christopher G. Baker$^1$}

\address{$^1$Centre for Engineered Quantum Systems, School of Mathematics and Physics, The University of Queensland, Australia}
\address{$^2$School of Engineering and Information Technology, UNSW Canberra, Canberra, Australian Capital Territory, Australia}
%ffiliation{$^1$ARC Centre for Engineered Quantum Systems, School of Mathematics and Physics, The University of Queensland, St Lucia, QLD 4072, Australia}
\ead{wbowen@physics.uq.edu.au}
\vspace{10pt}
\begin{indented}
\item[]April 15th, 2019
\end{indented}

\begin{abstract}
Vorticity in two-dimensional superfluids is subject to intense research efforts due to its role in quantum turbulence, dissipation and the BKT phase transition. Interaction of sound and vortices is of broad importance in Bose-Einstein condensates and superfluid helium \cite{gauthier_negative-temperature_2018, johnstone_order_2018, ellis_quantum_1993, parker_controlled_2004}. However, both the modelling of the vortex flow field and of its interaction with sound are complicated hydrodynamic problems, with analytic solutions only available in special cases. In this work, we develop methods to compute both the vortex and sound flow fields in an arbitrary two-dimensional domain. Further, we analyse the dispersive interaction of vortices with sound modes in a two-dimensional superfluid and develop a model that quantifies this interaction for any vortex distribution on any two-dimensional bounded domain, possibly non-simply connected, exploiting analogies with fluid dynamics of an ideal gas and electrostatics. As an example application we use this technique to propose an experiment that should be able to unambiguously detect single circulation quanta in a helium thin film. 

\end{abstract}

%Here we theoretically investigate the potential of  micrometer-scale whispering gallery mode cavities covered with thin films of superfluid $^4$He as optomechanical resonators. 
%Here we theoretically investigate the potential of  thin films of superfluid $^4$He covering micrometer-scale whispering gallery mode cavities as optomechanical resonators. 

\vspace{2pc}
\noindent{\it Keywords}: cavity optomechanics, superfluidity, superfluid helium films, optical resonators, third-sound, vortices, quantized circulation, persistent currents, 2D superfluids, vortex-sound coupling, Finite Element Modelling.
%\submitto{\NJP}
%\maketitle

\section{Introduction}
\label{sectionintroduction}

Superfluidity in two dimensions, first systematically investigated in the 70's in helium thin films\cite{bergman_hydrodynamics_1969, bergman_third_1971,berthold_superfluid_1977,bishop_study_1978}, has sparked major research efforts in recent years, culminating in the 2016 Nobel Prize, awarded for understanding the nature of superfluidity in two dimensions \cite{berezinskii_destruction_1971, berezinskii_destruction_1972, kosterlitz_ordering_1973}. The superfluid phase transition is native to a broad variety of physical systems, such as two-dimensional Bose-Einstein condensates\cite{hadzibabic_berezinskiikosterlitzthouless_2006,holzmann_superfluid_2007}, exciton-polariton condensates\cite{lagoudakis_quantized_2008,roumpos_single_2011}, and topological condensed matter systems\cite{qi_time-reversal-invariant_2009}.  Quantized vortices play a crucial role in these two-dimensional fluids, as their binding into pairs enables the emergence of long-range order and thereby transition to the superfluid phase. Understanding of vortex dynamics provides a pathway for controlling vortices with sound, imaging vortex distributions, understanding quantum turbulence, and engineering dynamical interactions between vortices and sound\cite{johnstone_order_2018,barenghi_introduction_2014,paoletti_quantum_2011,skrbek_quantum_2011}. 

Vortex dynamics in strongly interacting superfluids is of significance to a range of research fields: In topological condensed matter physics, it is responsible for the superfluid phase transition and the onset of dissipation \cite{berezinskii_destruction_1971, berezinskii_destruction_1972, kosterlitz_ordering_1973}; In astrophysics, the observed glitches in the rotation frequency of neutron stars are thought to result from vortex-unpinning events \cite{vortex_neutro_star_2016}; evidence for half quantum vortices (HQVs) has been found in superfluid $^3$He, where the HQVs in the A-phase of superfluid $^3$He are thought to host Majorana-fermions, bearing promise for fault tolerant topological quantum computing \cite{Leggett_review_1975,Volovik_book_1994,Nayak_review_2006}; vortices in $^3$He are interesting as analogues of exotic topological defects \cite{Volovik_book_1994,Makinen_HQV_2019} ---  the broken-symmetry-core vortex in superfluid $^3$He-B corresponding to Witten-strings \cite{Witten_superconduction_1985,Silaev_double_core_2015,Kondo_nonaxisymmetric_1991}; the HQV in the polar phase of superfluid $^3$He corresponding Alice-strings\cite{Alice_strings_1982}; the spin-mass-vortex in $^3$He-B, which has been proposed as an analogue for composite defects appearing in some grand unified theories of particle physics and even the standard model \cite{Kondo_spin_mass_vortex_1992,Eltsov_composite_defect_2000}. Therefore, the ability to determine the flow field induced by an arbitrary configuration of vortices, on an arbitrary and perhaps multiply-connected geometry is of broad importance, as is the ability to predict the strength of vortex-sound interaction.  

In the case of Bose-Einstein condensates, pressure- and temperature-waves constitute the relevant sound eigenmodes ({\it first-} and {\it second-sound}, respectively), while for helium thin films, both these modes of oscillation are suppressed due to the incompressibility of the fluid and the clamping of the normal fluid component. Surface excitations, so-called {\it third-sound} waves, become the primary form of sound wave \cite{atkins_third_1959}.  Recently, both observation of temperature-wave propagation in a two-dimensional Bose-Einstein condensate~\cite{ville_sound_2018}, and real-time measurement and control of {\it third sound} on a superfluid helium thin film~\cite{harris_laser_2016} have been demonstrated.

Quantifying vortex flow fields and their interaction with sound waves has sparked substantial research efforts. Ellis {\it et al.} \cite{ellis_quantum_1993,wilson_vortex_1995,ellis_excitation_1998} electrically excited third-sound modes to swirl up an ensemble of vortices in a helium thin film --- however, despite elaborate mathematical analysis \cite{wilson_swirling_1998}, their modelling of vortex-sound interaction was limited to simple, centered vortex ensembles on a circular resonator. In the case of BECs, a plethora of analyses for vortex-sound interaction has been performed \cite{PhysRevA.58.3168,PhysRevLett.81.1754,parker_controlled_2004,White_vortices_2014,Lucas_sound_2014,Berloff_interactions_2004,Sinha_semiclassical_1997,PhysRevA.56.587,PhysRevB.74.224503} and the dispersive interaction has been quantified for centered vortices in simple trap geometries \cite{PhysRevLett.81.1754,PhysRevA.58.3168}.

In this work, we model vortex flow fields and the interaction of sound modes with vortices, in a two-dimensional superfluid by exploiting analogies with other areas of physics. We map vortex dynamics onto electrostatics, and superfluid hydrodynamics onto fluid dynamics of an ideal gas. This allows us to draw on technically mature finite-element-modelling (FEM) tools available for these fields. We show how the interactions of sound and the flow field of arbitrary vortex distributions can be computed using these tools on any two-dimensional, not necessarily simply connected, domain. Thus, our work provides a theoretical framework for controlling and imaging vortices, and for engineering a dynamical interaction between sound and vortices.

As an example, we discuss the interaction of sound Bessel modes on a disk-shaped domain with quantized vortices, which is relevant for a number of experiments on superfluid thin films\cite{harris_laser_2016,mcauslan_microphotonic_2016,sachkou_dynamics_2018,ellis_quantum_1993} and two-dimensional
Bose-Einstein-condensates~\cite{matthews_vortices_1999, weiler_spontaneous_2008,ville_sound_2018}.  The interaction induces splitting between otherwise degenerate sound modes \cite{ellis_quantum_1993,PhysRevLett.81.1754,PhysRevA.58.3168}. We show how the vortex number can be extracted from experimental measurement of the splitting. Further, we present a perturbative analytic model, approximating the FEM-simulation for circular geometries, which offers some intuitive insight on the vortex-sound coupling mechanism.

Lastly, we focus our analysis to the prospect of detecting quantized circulation in helium thin films. While vortices in Bose-Einstein condensates can be visualized by optical snapshots \cite{Freilich1182}, and pinned vortices in exciton-polariton condensates can be visualized by optical interferometry, no such direct observation technique exists for helium thin films: the vortex core is an {\AA}ngstr\"{o}m-size perturbation on an ultra-thin film of transparent liquid, whose flow does not interact dissipatively with the environment.  Measurements on helium are important because, unlike exciton-polariton condensates and most Bose-Einstein condensates, the atoms in superfluid helium are strongly interacting, introducing dynamics that can not be modelled through the Gross-Pitaevskii equation, and are not fully understood \cite{tilley_superfluidity_1990}. 

We finally propose an experiment where discrete steps due to an increase or decrease in the number of circulation quanta could be observed for the case of a superfluid helium film. We suggest a geometry where vortex pinning around an engineered topological defect leads to experimentally observable quantized steps. Drawing on the finite element model, we discuss how, in this geometry, the interaction with sound can be maximized, so that these steps could be clearly resolved. This would enable the first direct detection of quantized circulation in  two-dimensional superfluid helium.

\section{Sound-vortex interaction and their analogues}
\subsection{Sound in two-dimensional superfluids}
\label{Thirdsoundtheorysection}

Superfluid hydrodynamics is generally described by the continuity equation\cite{ellis_quantum_1993}: 
\begin{equation}
\frac{d\rho}{dt}=-\vec{\nabla} \cdot (\rho\vec{v}),
\label{tscontinuity}
\end{equation}
which derives from mass conservation, and the Euler equation:
\begin{equation}
\frac{d\vec{v}}{dt}+(\vec{v}\cdot\vec{\nabla})\vec{v}=- g\vec{\nabla}\rho+C,
\label{tsEuler}
\end{equation}
which derives from Newton's second law (momentum conservation). In the case of superfluid helium thin films,  $\vec{v}$ is the superfluid flow velocity, $\rho\rightarrow h$ is the film height, $g=\frac{3\alpha_{\mathrm{vdw}}}{h^4}$ is the linearized Van-der-Waals acceleration\cite{atkins_third_1959, tilley_superfluidity_1990}, and $C=0$. In the case of Bose-Einstein-condensate hydrodynamics in the Thomas-Fermi-limit at zero temperature \cite{schuck_thomas-fermi_2000,pitaevskiiBEC}, $\vec{v}$ is again the flow velocity, $\rho$ is the density, $g\rightarrow g_{\mathrm{BEC}}/M^2 $ describes the coupling strength, where $M$ is the mass of an individual atom contributing to the condensate, $g_{\mathrm{BEC}}$ the atom-atom coupling, and $C=-\vec{\nabla}U/M$ describes the trapping, with $U$ being the extended potential\cite{schuck_thomas-fermi_2000} (see Appendix, table \ref{tab:acoustics table}). 

\begin{figure}[h]
\centering
\includegraphics[width=\textwidth]{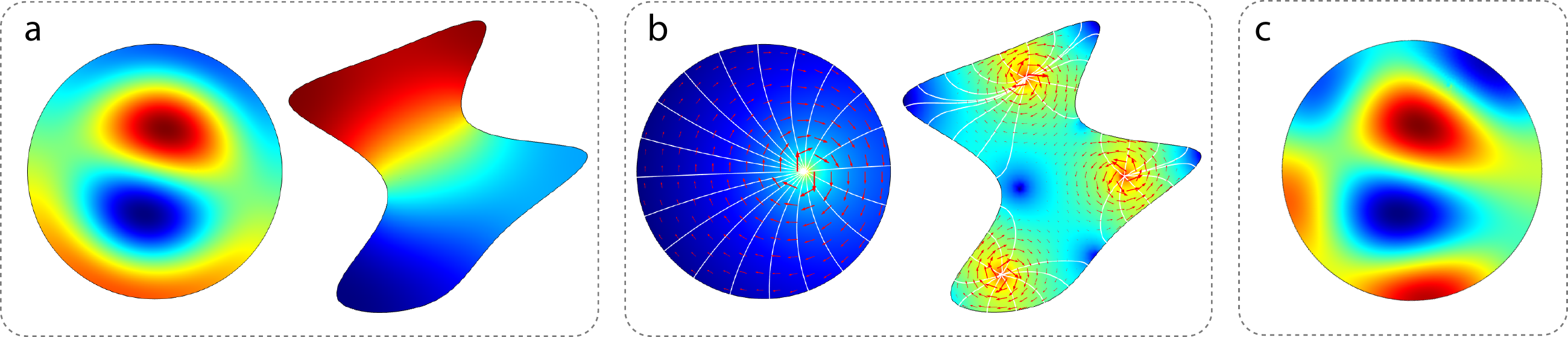}
\caption{(a) Finite Element Method (FEM) modelling of the sound eigenmodes existing within a circular (left) and an arbitrarily shaped domain (right), with free boundary condition (see \ref{sec:boundary conditions}). Left: Bessel ($m$=1; $n$=2) mode. Right: Lowest-frequency eigenmode of the geometry. Color code represents the magnitude of the displacement. (b) FEM  modelling of the flow field generated by point vortices within the domains shown in (a). Left: simple case of the flow field of an off-centered clockwise (CW) point vortex in a circular domain (see \ref{sectionanalyticalpointvortex}). Right: flow field due to two CW and one counterclockwise (CCW) point vortices. Surface color-code and red arrows show the vortex flow velocity, in log scale. White lines represent the streamlines of the unrotated electric displacement field $\vec{D}$, which are potential lines for the superfluid flow.  (c) New `deformed' ($m$=1; $n$=2)  eigenmode of the circular geometry in the presence of the background flow due to a large number of off-centered vortices located at a point with radial offset of 0.7 R, where R is the resonator radius.}
\label{Figure1fig}
\end{figure}

We assume small perturbations in the density (BEC) or film height (helium), $\eta$, from an equilibrium $\rho_0$, $\rho(\vec{r},t)=\rho_0+\eta(\vec{r},t)$ with $\eta \ll \rho_0$. Eqs (\ref{tscontinuity}) and (\ref{tsEuler}) respectively become:

\begin{equation}
\dot{\eta}=-\rho_0\vec{\nabla} \cdot \vec{v} - \vec{v}\,\vec{\nabla}\eta,
\label{tslincontinuity}
\end{equation}
and
\begin{equation}
\dot{\vec{v}}+(\vec{v}\cdot\vec{\nabla})\vec{v}=- g \vec{\nabla} \eta+C,
\label{tslinEuler}
\end{equation}
where we used the chain rule on the r.h.s. in Eq. (\ref{tsEuler}). Substituting $g \rightarrow c^2=\gamma RT/\rho_0 $, where $R$ is the specific gas constant, $T$ the gas temperature and $c$ the speed of sound, we find that the above equations are the linearized Euler and continuity equations, describing small amplitude sound waves in an ideal gas in the isentropic limit \cite{rienstra_introduction_2004} (see \ref{seclinac} for derivation). This allows us to model them using the {\it Aeroacoustics $\rightarrow$ Linearized Euler, Frequency Domain (lef) module} in COMSOL, with appropriate boundary conditions (see \ref{sec:boundary conditions}). This provides the sound eigenmodes for an arbitrary bounded geometry. Examples of eigenmodes on a circular and an irregularly shaped two-dimensional resonator with  free (`{\it Neumann}') boundaries are shown in Fig.~\ref{Figure1fig}(a) \cite{sachkou_dynamics_2018} (see \ref{sec:boundary conditions}). In this work, we analyse vortices and sound modes in two-dimensional domains, as this applies to experiments \cite{ellis_quantum_1993,gauthier_negative-temperature_2018,sachkou_dynamics_2018} and allows a qualitative discussion of the interaction. However, by choosing the appropriate dimension in the FEM-simulation, the model could be generalized to three dimensions. The finite-element-method allows us to add a background flow field, corresponding to the flow generated by quantized vortices, and find the new sound eigenmodes in the presence of that background flow.

\subsection{Quantized vortices}
\label{Quantizedvorticestheorysection}

A single vortex is described by a quantized circulation around a loop encompassing the vortex core\cite{ambegaokar_dynamics_1980,donnelly_quantized_1991} :

\begin{equation}
\oint \vec{v}_v \cdot d\vec{l} = \kappa.
\label{eqvortex}
\end{equation}
Here, $\kappa=2\pi\hbar/M$ is the circulation quantum. This ensures that the phase acquired by the wave function upon propagation around any loop encompassing the vortex core equals $2\pi$.  $\vec{v}_v$ denotes the vortex-induced velocity field. For the simple case of a point vortex on a plane, the solution is:
\begin{equation}
 \vec{v}_v(r)=\frac{\kappa}{2\pi r}\hat{e}_\theta,
\end{equation}
where $\hat{e}_\theta$ is the unit vector in the tangential direction and $r$ the distance from the vortex core. In this work, we describe the quasi-static regime where the motion of vortices during a sound oscillation period is negligible. This is valid in the limit of pinned vortices \cite{ellis_quantum_1993, ellis_excitation_1998, hoffmann_measurements_2004} or low vortex densities where the velocity of the flow field due to neighboring and image vortices is significantly less than the speed of sound. For Bose-Einstein condensates at zero temperature, the sound velocity $c=\hbar/M\xi$, where $\xi$ is the healing length, equals the Landau critical velocity \cite{pitaevskiiBEC}. Therefore the quasi-static approximation in Bose-Einstein condensates is valid if the separation between neighbouring vortex cores is significantly larger than their core diameter. This condition is typically fulfilled, and the sound velocity, with typical values of a few mm/s \cite{Ketterle_BEC,Ketterle_BEC_Erratum} is significantly higher than the background flow velocity caused by a typical ensemble of vortices \cite{gauthier_negative-temperature_2018,johnstone_order_2018}\footnote{The orbit period T for a single vortex offset from the center of a circular resonator of radius R by a distance $x$ is $T=\frac{4\pi^2}{\kappa}\left( R^2-x^2 \right)$. For superfluid helium thin film resonators with $R\simeq 10^{-5}$ m, this corresponds to typical Hz orbit frequencies compared to typical $10^5$ Hz third-sound frequencies \cite{harris_laser_2016}.}. 

In comparison to Eq. (\ref{eqvortex}), Gauss's law of electrostatics in two dimensions reads:

\begin{equation}
\oint \vec{D}\cdot d\vec{n}=Q,
\end{equation}
where $\vec{D}$ is the electric displacement field and $Q$ the line charge. By rotating the electric displacement field and replacing it with the flow field $\vec{v}_v$

\begin{equation}
\left(
\begin{array}{c}
D_x \\
 D_y \\
\end{array}
\right)\rightarrow\left(
\begin{array}{c}
v_{v\,y} \\
 -v_{v\,x} \\
\end{array}
\right),
\label{EqpermutationDgoestov}
\end{equation}
and substituting $Q \rightarrow \kappa$, we retrieve the quantized circulation of the vortex defined in Eq. (\ref{eqvortex}), where $\vec{v}_v=v_{v\,x}\, \hat{e}_x+v_{v\,y}\, \hat{e}_y$. This provides the analogy between electrostatics and vortex flow \cite{ambegaokar_dynamics_1980}. A point charge is a source of divergence (source/sink) for the electric displacement field $\vec{D}$. As is known from potential flow theory, upon the permutation shown in Eq. (\ref{EqpermutationDgoestov}), a point charge becomes a source of quantized circulation. We model these equations using the {\it Electrostatics(es) module} of COMSOL, which allows us to determine the vortex flow field on any two-dimensional geometry. Examples for vortex flow fields on a circular and on an irregular geometry are shown in Fig. \ref{Figure1fig}(b). Depending on the number of vortices, their positions, and the resonator geometry, the sound eigenmode shape may be significantly altered due to the presence of the vortices. Such an example is shown in Fig. \ref{Figure1fig}(c). Similarly, quantized circulation around a macroscopic topological defect in a multiply-connected domain can be modelled as shown in section \ref{sectionsinglequantadetection}.

\begin{figure}
\centering
\includegraphics[width=.7\textwidth]{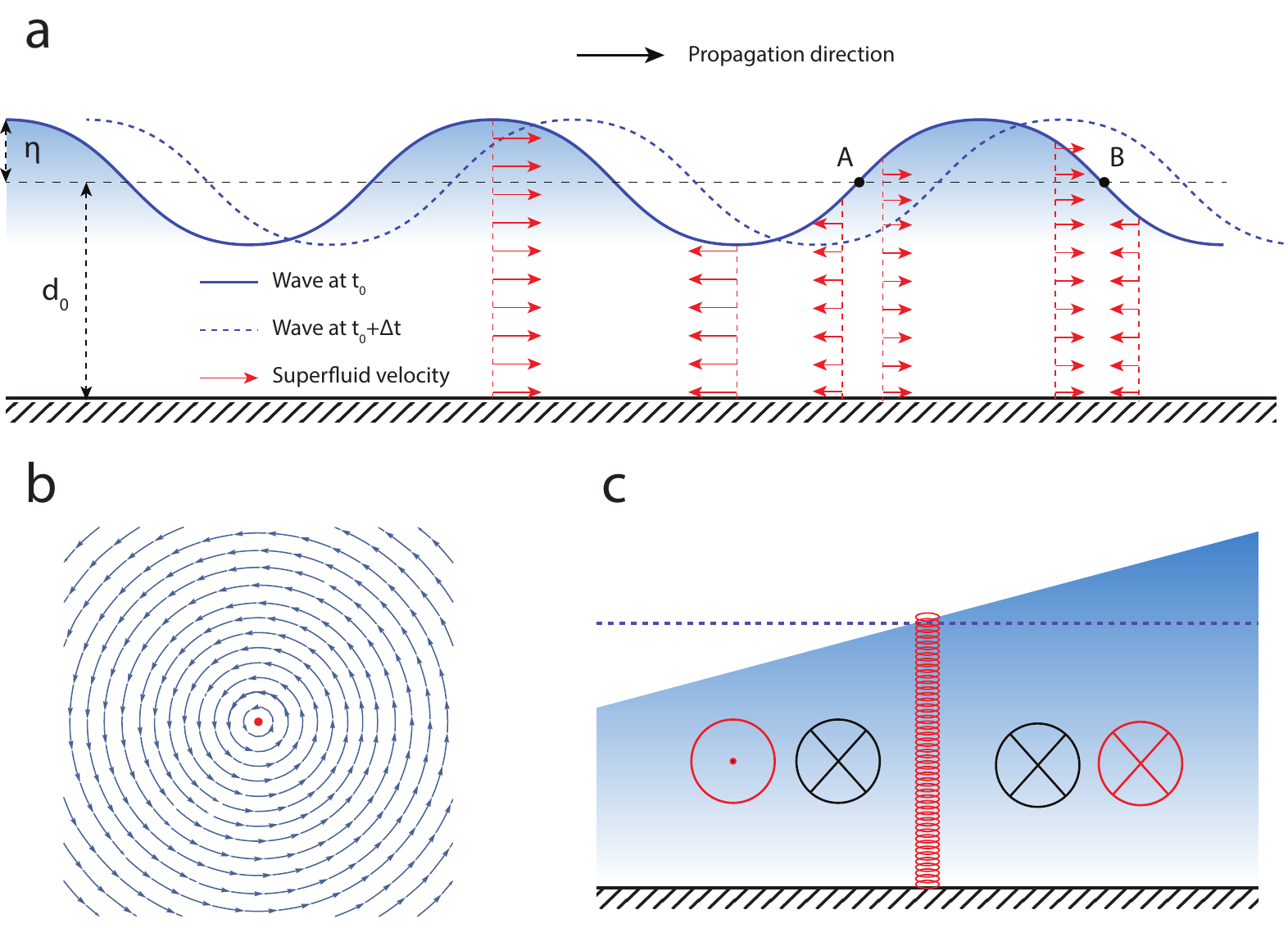}
\caption{(a) Illustration of the flow profile of sound in superfluid. (b) Streamlines of a 2D point vortex (red dot) in the plane. (c) Vortex flow (red) and sound flow (black). When sound and vortex flow fields are confined inside a resonator geometry (see Fig. (\ref{Figureanalyticalexplanationfig})), interference effects arise. Destructive interference (left) and constructive interference (right), in the presence of a mode-induced height gradient, causes an interaction between vortex and sound. The vertical axes in (a) and (c) can refer to density in a Bose-Einstein condensate, or the film height in a superfluid helium thin film. }
\label{Figure0fig}
\end{figure}

\subsection{Sound-vortex interaction}
\label{soundvortexinteraction}

We can understand the interaction of sound waves and vortices through the change in the kinetic energy of a sound wave caused by addition/subtraction of a vortex. The sound modes are orthogonal to vortex flow fields, which are fully defined by the rotation around the vortex core (see Fig. \ref{Figure0fig} (b)). Therefore, the overlap of vortex- and sound velocity fields is zero:

\begin{equation}
\int_A \vec{v}_v\cdot \vec{v}_s\, dA=0,
\end{equation}
where $A$ is the area of the domain and $\vec{v}_s$ is the two-dimensional sound velocity distribution. This appears to suggest that there is no coupling between the two flow fields. However, interaction arises due to the change in film height (helium) or density (BEC) associated with the sound wave. The interaction manifests in a splitting of sound resonances due to the presence of vortices, and thus constitutes a dispersive (frequency-shifting) interaction \cite{Aspelmeyer}. An example is shown in Fig. \ref{Figure0fig}(c), where, due to increased density/height on one side of the vortex, the increased kinetic energy due to velocity addition on the right side of the vortex is not fully compensated by the reduction in kinetic energy due to velocity subtraction on the left, resulting in a net increase in energy due to the interaction.

%to first order, one would expect no coupling between the two. However, locally, third sound flow and vortex flow add up to change the kinetic energy of the fluid. One way an interaction between the two arises is, when the flow fields add up %constructively on one side of the resonator, and destructively on the other side, with a mode-induced density/height gradient between the sides (see Fig. \ref{Figure0fig}(c)).

We use Eqs (\ref{tslincontinuity}) and (\ref{tslinEuler}) to calculate the new eigenmodes of sound in the presence of the time-independent background flow $\vec{v}_v$, assuming a stationary configuration of vortices, with flow field $\vec{v}_v(\vec{r})$. In this case, the total flow velocity is $\vec{v}(\vec{r},t)=\vec{v}_v(\vec{r})+ \vec{v}_s(\vec{r},t)$, where $\vec{v}_s$ is the flow field associated with the sound eigenmode.  The interaction of persistent currents with sound modes has been quantified for simple, centered vortex distributions on a disk-shaped resonator in superfluid helium thin films\cite{ellis_quantum_1993,wilson_swirling_1998} and for centered vortices in different trap shapes in Bose-Einstein-condensates \cite{PhysRevLett.81.1754,PhysRevA.58.3168}, but until now there has not been a consistent approach for modelling of a non-trivial vortex distribution in a non-trivial resonator shape, or for multiply connected domains. We confirm the accuracy of this finite element based approach through the comparison with an analytical expression derived for third-sound-vortex interactions on a circular resonator (see \ref{sec: analytical expression}).

\section{Results}
\label{sectionresults}

\begin{figure}
\centering
\includegraphics[width=\textwidth]{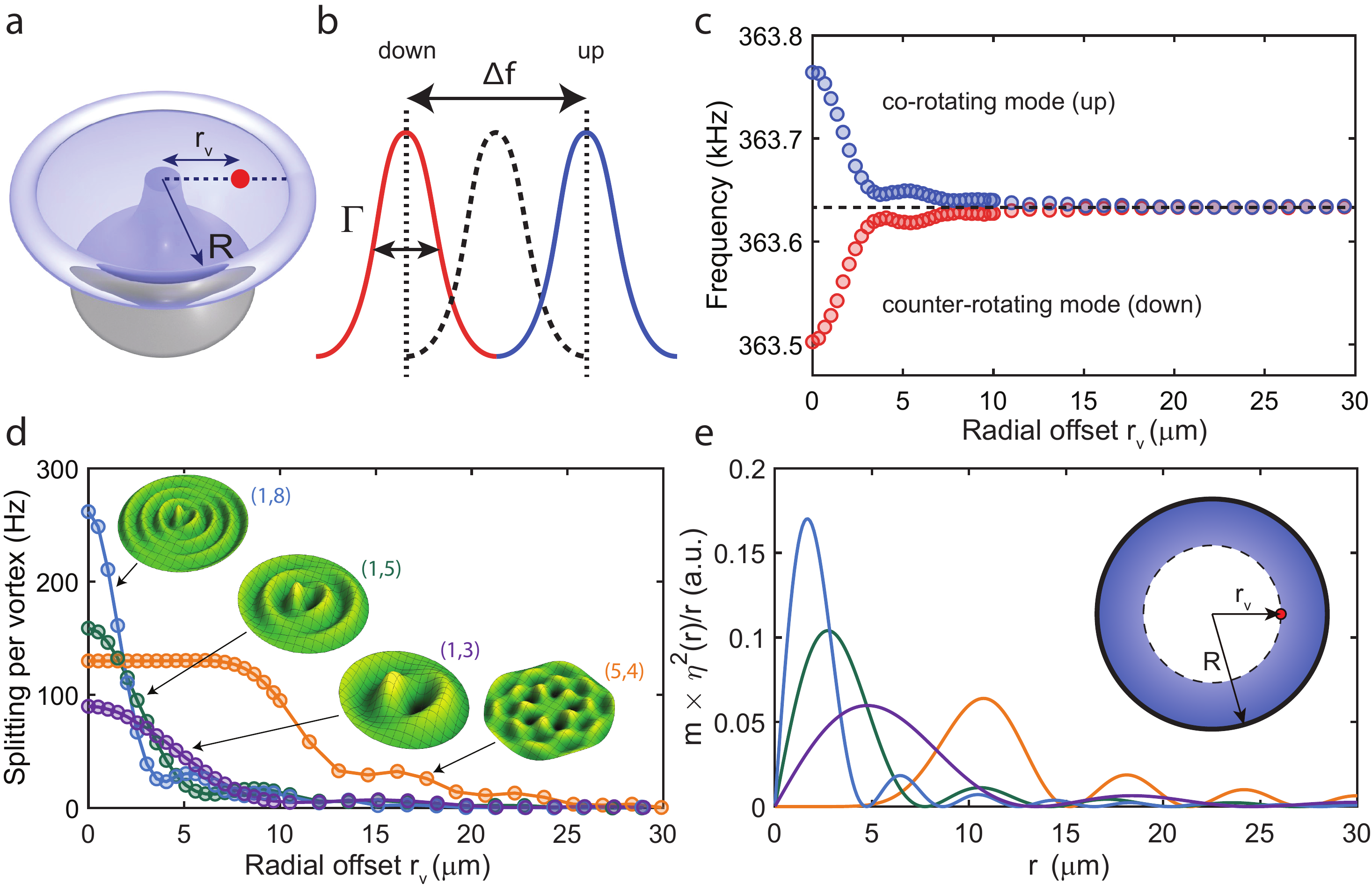}
\caption{ (a) Schematic of a microtoroidal resonator of radius $R=30 ~\mu$m covered with a superfluid helium thin film, with one quantized vortex offset from the disk origin. The red dot indicates the vortex core. (b) Illustrated frequency splitting of a Bessel mode due to the presence of a vortex. (c) Frequency of the co- and counter-rotating (1,8)-Bessel mode, with one quantized vortex on the resonator, as a function of radial vortex position. (d)  Frequency splitting dependence of (1,3), (1,5), (1,8), and (5,4) third-sound modes on the radial offset of the vortex from the disk origin. Spatial profiles of the modes are shown as insets. (e) Displacement amplitudes of modes shown in (d) as a function of radius. Inset: annular region cut out by a circle, with its radius defined by the vortex position - whose interaction with the vortex accounts for the majority of the splitting (see text). All plots use free boundary condition for the modes; each dot in (c) and (d) represents the result of a FEM-simulation.}
\label{Figure2fig}
\end{figure}

To give an experimentally relevant example, in the following we study the interaction of a quantized vortex with sound modes in a disk-shaped resonator with a free (`{\it Neumann}') boundary condition (see \ref{sec:boundary conditions}). This analysis is applicable to geometries used in superfluid helium experiments\cite{harris_laser_2016,mcauslan_microphotonic_2016,sachkou_dynamics_2018} with a microtoroidal optomechanical resonator of $R \sim 30 ~\mu$m radius  (see Fig. \ref{Figure2fig}(a)), those of refs. \cite{ellis_observation_1989, ellis_quantum_1993, schechter_observation_1998, hoffmann_measurements_2004}, and also experiments with two-dimensional Bose-Einstein-condensates, which are  confined by a hard-walled trap \cite{gauthier_negative-temperature_2018}. 

Regarding the experimental readout of sound modes, in experiments with helium thin films, the Brownian motion of third sound waves, even at millikelvin temperatures, is high enough to be resolved experimentally in real time \cite{harris_laser_2016}. Alternatively, the amplitude of third sound can be tuned by laser heating or cooling \cite{harris_laser_2016}, amplified by optical absorption heating \cite{mcauslan_microphotonic_2016}, or electrically excited \cite{ellis_quantum_1993}. For Bose-Einstein condensates, collective modes can be excited by perturbing the condensate, locally exceeding the critical velocity \cite{Pethick_BEC}.

Solutions of the wave equation on a circular resonator are Bessel modes of the first kind. They are fully quantified by their azimuthal ($m \geq 0 $) and radial ($n \geq 1$) node counts. Modes with $m\neq 0$ can be decomposed into opposite direction travelling waves, that, in the absence of circulation, are degenerate. The presence of a vortex lifts the degeneracy, shifts the mode co-rotating with the vortex flow to a higher frequency, and the counter-rotating mode to a lower frequency (see Fig. \ref{Figure2fig}(b)). This frequency splitting $\Delta f$ is experimentally resolved if it is larger than the decay rate $\Gamma$ of the Bessel mode. As the radial distance of the vortex from the center of the disk, $r_v$, increases, the splitting reduces. Firstly, this results from the lower total energy of the vortex flow field as the vortex core approaches the boundary. Secondly, from symmetry arguments the overlap between sound and vortex flow fields is maximized for a centered vortex. The radial dependence of the mode shifts is plotted for the ($m$,$n$)=(1,8) free-boundary-condition Bessel mode in Fig. \ref{Figure2fig}(c). Each point corresponds to one result of the FEM-simulation, as the vortex is stepped outwards from the center. As the vortex reaches the outer boundary, the splitting vanishes.

We then compare the results from the FEM-model to the perturbative analytical approach derived in \ref{sec: analytical expression}. We find that the two approaches agree reasonably well, with the discrepancy always less that 10\% of the maximal splitting at $r_v=0$ (see Fig. (\ref{Figure7fig})). We ascribe the difference to a vortex-induced change in third-sound mode shape due to the nonlinear $(\vec{v}\cdot\vec{\nabla})\vec{v}$ term in Eq. (\ref{tslinEuler}):

\begin{equation}
(\vec{v}\cdot\vec{\nabla})\vec{v}=(\vec{v}_v\cdot\vec{\nabla})\vec{v}_v+(\vec{v}_s\cdot\vec{\nabla})\vec{v}_s+(\vec{v}_v\cdot\vec{\nabla})\vec{v}_s+(\vec{v}_s\cdot\vec{\nabla})\vec{v}_v.
\label{eqvgradv}
\end{equation}
The analytic perturbation theory neglects all of these terms. The FEM model neglects only $(\vec{v}_s\cdot\vec{\nabla})\vec{v}_s$, which is a requirement to obtain eigenmodes for the sound waves.
This approximation is justified in the small sound amplitude limit $\eta \ll \rho_0$, where the sound-induced superfluid flow velocity $v_s$ (see Eq. (\ref{Eqv3})) is small compared to the vortex flow velocity $v_v$.

Fig. \ref{Figure2fig}(d) shows the splitting per vortex as a function of vortex radial position for four different Bessel modes. Critically, the presence of a vortex affects each Bessel mode in a unique fashion. Leveraging this unique fingerprint, the work presented here enabled both the number and the spatial distribution of vortices in a cluster to be extracted independently, by tracking several sound modes simultaneously \cite{sachkou_dynamics_2018}. 

One conceptual result of the perturbative analysis is an expression for the frequency splitting that depends only on the profile of the Bessel mode and the radial position of the vortex, and is independent of the details of the vortex flow field. In the case of a single vortex the expression is:

\begin{equation}
\centering
\Delta f=\frac{\kappa\,m}{2\pi^2}\,\frac{\int_{r_v}^R\frac{\rmd r}{r}\,\eta^2\left(r\right)}{\int_{\mathrm{0}}^{R} \rmd r \,r\,\eta^2\left(r\right)},
\label{eqsplittingmaintext}
\end{equation}
where $m$ is the Bessel mode azimuthal number, $r_v$ is the radial position of the vortex, and $\eta\left(r\right)=J_m\left( \zeta_{m,n} \, \frac{r}{R}\right)$ is the radial displacement profile of the Bessel mode. $J_m$ is the Bessel function of the first kind of order $m$, and $m$ and $n$ are respectively the azimuthal and radial mode orders, $\zeta_{m,n}$ is a frequency parameter depending on the mode order and the boundary conditions \cite{baker_theoretical_2016}. As vortex flow fields are subject to linear superposition, we obtain the total splitting simply by adding contributions from single vortices: $\Delta f_{\rm total}= \sum_i\Delta f_i$. An interpretation of this result is that the splitting introduced by a vortex at position $r_v$ is equal to the interaction energy between a centered vortex ($r_v=0)$ and a sound wave in the region of the disk with radius greater that $r_v$. So, in some sense, only the fraction of the sound wave at radius larger than the radial position of the vortex core contributes to the splitting. This explains why the splitting per vortex drops rapidly with $r_v$ for low $m$, high $n$ order sound modes whose kinetic energy is located close to the center of the disk, while the splitting is sustained at higher radii for higher $m$, lower $n$ order modes which are more radially extended.

\section{Requirements for detection of single vortices/circulation quanta}
\label{sectionsinglequantadetection}

In this section, we investigate the feasibility of observing the quantization of circulation in two-dimensional superfluid helium due to the shift in sound frequencies induced by the addition/subtraction of a single vortex. Remarkably, while quantized vortices are central to the behavior of two-dimensional superfluids, they have yet to be directly observed in two-dimensional helium. 
The experimental challenge is significant: 
the normal-fluid core of a vortex in superfluid helium-4 is roughly one Angstr\"{o}m in diameter  \cite{tilley_superfluidity_1990},
the thickness of a superfluid helium film is typically less than 20~nm, and
the refractive index of liquid helium is close to that of vacuum ($n_{\rm He} \approx 1.029$). Combined, these characteristics prevent direct optical imaging, as can be performed in Bose-Einstein condensates~\cite{weiler_spontaneous_2008, wilson_situ_2015}. %cite{weiler_spontaneous_2008}. 
In bulk helium, many  imaging techniques have relied on the use of some kind of tracer particle \cite{yarmchuk_observation_1979, bewley_superfluid_2006, fonda_direct_2014}, such as, for instance, micrometer-sized frozen hydrogen crystals. These scatter light and are pulled in to the vortex core, enabling, for example, the recent observation of Kelvin waves \cite{fonda_direct_2014} in bulk. Naturally, such an approach is significantly more difficult in two-dimensional films due to their few-nanometre thickness. 

In order for the  vortex-induced frequency splitting $\Delta f$ (Eq.(\ref{eqsplittingmaintext}))  experienced by a third-sound 
wave \cite{ellis_observation_1989, ellis_quantum_1993} to reveal quantized steps, several challenges must be addressed.\\
First, in order to be resolvable, the splitting should be larger than the linewidth of the third-sound resonances, $\Delta f\gtrsim \Gamma$, as shown in Figure \ref{Figure2fig}(a). Second, any motion of vortices on the resonator surface, as we experimentally observe elsewhere \cite{sachkou_dynamics_2018}, will lead to a continuous evolution of the splitting due to the continuous nature of the splitting function $\Delta f(r_v)$, see Figure \ref{Figure2fig}(c), which may mask the quantized nature of the circulation. %stand in the way of the observation of discrete steps.

The first challenge can be met by engineering devices that are sufficiently small, which maximizes $\Delta f$ by increasing the vortex-sound coupling, and by controlling dissipation in these devices in order to reduce $\Gamma$. This can be achieved, for instance, by engineering a smooth resonator that is decoupled from its environment by a small connection point \cite{harris_laser_2016}, or through careful choice of the resonator substrate material \cite{Davis_ultralow_2017}. Indeed, $\Delta f\sim \Gamma$ has been recently reported experimentally using microscale on chip optical cavities \cite{sachkou_dynamics_2018}. A solution to the second challenge is to constrain the position of the circulation around a macroscopic topological defect engineered on the surface of the resonator. For instance, if we replace the topological defect naturally formed by the normal fluid core (of radius $a_0$) of a superfluid vortex by a microfabricated hole of radius $R\gg a_0$, the maximal velocity due to the quantized circulation becomes $\frac{\kappa}{2\pi R}\ll \frac{\kappa}{2\pi a0}$ (see Eq. (\ref{eqvortex})). This effectively clips the high velocity region of the flow and is thus energetically favourable. The circulation will then preferentially accumulate around this manufactured defect, up to large values of $\kappa=h/m_{\rm He}$, as has been observed in the spinning up of bulk helium in an annular container \cite{bendt_superfluid_1967, fetter_low-lying_1967}.
The quantization of the circulation then manifests as quantized values of the splitting experienced by third-sound modes confined to the surface of the resonator.

\begin{figure}
\centering
\includegraphics[width=\textwidth]{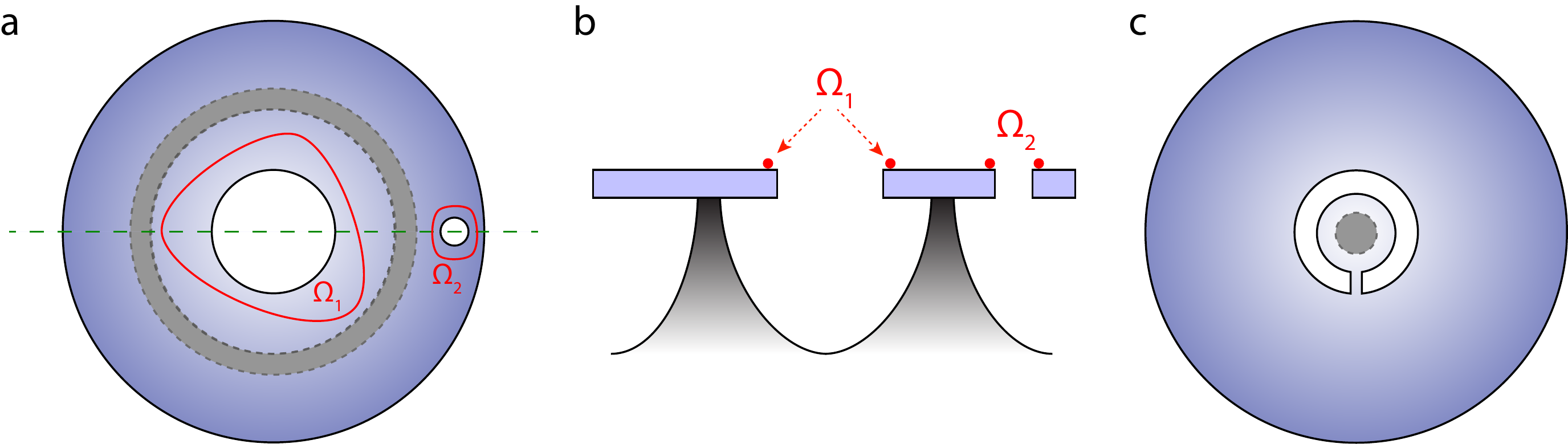}
\caption{ Topological defect in a disk-shaped resonator geometry. (a) Top-view of an annular-shaped superfluid optomechanical resonator \cite{baker_theoretical_2016}. Blue shade represents the disk, while the dashed grey region symbolizes the device's pedestal. Red contours $\Omega_1$ and $\Omega_2$ represent closed loops around holes in the resonator. (b) Cut-view through the dashed green line in (a), illustrating how contour $\Omega_1$ can be continuously deformed and collapsed, while contour $\Omega_2$ cannot and encloses therefore a real topological defect. (c) Single-spoked annular disk geometry\cite{baker_high_2016}, whose central hole is topologically identical to that enclosed by contour $\Omega_2$.}
\label{Figure3fig}
\end{figure}

This approach is in essence a two-dimensional analogue of Vinen's experimental technique for the first observation of circulation quanta in bulk helium \cite{vinen_detection_1961, whitmore_observation_1968}, where circulation trapped around a vibrating wire lifted the degeneracy between the wire's normal modes of vibration. In Fig.(\ref{Figure3fig}), we propose a practical realization of such a device based on a circular whispering-gallery-mode geometry, as used in our previous work \cite{harris_laser_2016, mcauslan_microphotonic_2016}.  We utilise the FEM-simulation to design a domain that maximises the splitting $\Delta f$, consisting of a single-spoked annular geometry \cite{baker_high_2016} (see Fig. \ref{Figure3fig}(c)).

\begin{figure}
\centering
\includegraphics[width=\textwidth]{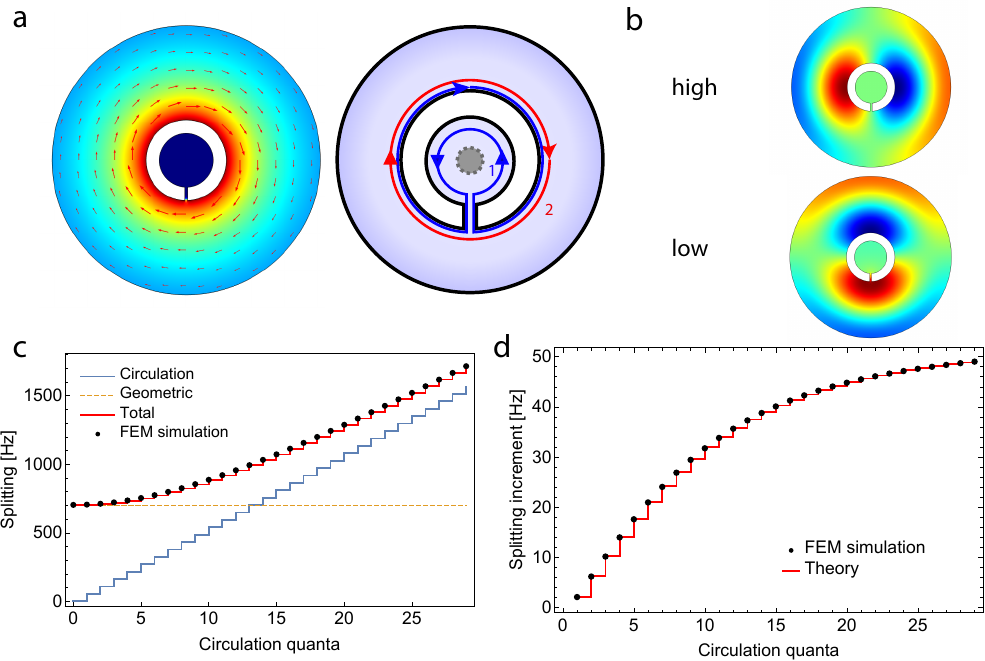}
\caption{ (a) Left: Superfluid flow field resulting from quantized circulation, pinned on the topological defect, with no additional vortices on the domain (Red arrows). Color code represents the magnitude of the velocity (red: fast; blue: slow). Right: contours 1 (blue) and 2 (red) both enclose the central defect and therefore have identical circulation (Eq.(\ref{eqvortex})). Dimensions of the resonator used for simulation: outer radius = 20~$\mu$m; inner radius = 4~$\mu$m; slot width = 2~$\mu$m;  spoke width = 0.5~$\mu$m. (b) Non-degenerate eigenmodes of the spoked resonator, split by the presence of the spoke, with free boundary conditions at both boundaries. Color code represents the magnitude of the surface displacement. (c) Contribution to the total mode splitting (red, $s_{\mathrm{total}}=\sqrt{s_{\mathrm{circ}}^2+s_{\mathrm{geo}}^2}$) from geometric (dashed orange, $s_{\mathrm{geo}}=700$~Hz) and circulation (blue,  $s_{\mathrm{circ}}=N\times s_{\mathrm{circ,0}}=N\times 54$ Hz) contributions, as a function of the number of circulation quanta present around the topological defect. Note that the number of circulation quanta considered here is consistent with vortex numbers observed in experiments with helium-coated microresonators \cite{sachkou_dynamics_2018}.  (d) Splitting increment per added circulation quantum.
 }
\label{Figure4fig}
\end{figure}

Next we calculate the superfluid flow field resulting from quantized circulation about the central topological defect, as shown in Fig. \ref{Figure3fig}(c). For clarity, we assume that this pinned circulation is the only source of circulation on the structure, i.e. there are no vortex cores on the domain. This is calculated through FEM simulation using the `\textit{floating potential}' boundary condition for the inner boundary, which enforces both the prescribed  circulation strength and parallelism of the flow to the boundary (see \ref{sec:boundary conditions} for more details). The results are shown in Figure \ref{Figure4fig}(a). The flow is essentially confined to the outer annulus, with negligible flow up and down the spoke and in the central disk. This can be understood by considering the closed contours 1 and 2 which both enclose the central hole. The circulation around both contours must therefore be equal (see Eq. (\ref{eqvortex})), implying negligible additional circulation along the extra path contained in contour 1.

Figure \ref{Figure4fig}(b) shows an example of a third-sound mode of this spoked resonator 
(which becomes the (1, 2) eigenmode of a circular resonator as the central hole gets vanishingly small). The presence of the spoke, which connects the annular outer ring to the device pedestal, lifts the degeneracy between the two normal modes, even in the absence of circulation. The mode that has a stronger interaction with the spoke (bottom) experiences an effectively larger resonator and therefore has a lower resonance frequency.

In Figure \ref{Figure4fig}(c), we show how this native \textit{geometric} splitting \cite{ellis_observation_1989} affects the third-sound mode splitting as a function of the circulation around the central hole. Each black dot represents a finite element simulation of the  splitting between the high and low frequency eigenmodes shown in Figure \ref{Figure4fig}(b), as a function of the number of circulation quanta around the central defect. This total splitting can be well reproduced through an analytical expression of the form \cite{ellis_excitation_1998}:
\begin{equation}
s_{\mathrm{total}}=\sqrt{s_{\mathrm{circ}}^2+s_{\mathrm{geo}}^2}.
\label{Eqgeocircsplitting}
\end{equation}
where $s_{\mathrm{geo}}=700$ Hz is the native geometric splitting for this device (dashed orange line), $s_{\mathrm{circ}}=N \times s_{\mathrm{circ,0}}$ the total circulation-induced splitting with N being the number of circulation quanta and $s_{\mathrm{circ,0}}$ is the splitting per quantum. The solid red line represents $s_{\mathrm{total}}$, as given by Eq. (\ref{Eqgeocircsplitting}). In Figure \ref{Figure4fig}(d), we plot the experimentally relevant parameter, which is the splitting increment due to each additional circulation quantum, as a function of the number of circulation quanta already present around the defect. This shows that the  geometric splitting due to the spoke (or any unwanted deviation from circularity) will mask the influence of the circulation-induced splitting for small values of the circulation quanta, and reduce the visibility of the steps. For larger values of the circulation, the size of the steps will asymptote towards the value $s_{\mathrm{circ,0}}=54$~Hz, as the appropriate normal mode basis gradually shifts from orthogonal standing waves to counter-propagating waves. Such a large quantized circulation can be experimentally achieved by creating a strong superfluid flow, locally exceeding the critical velocity --- for instance by local evaporation of superfluid or by strong driving of third-sound modes \cite{harris_laser_2016,mcauslan_microphotonic_2016,sachkou_dynamics_2018}. Alternatively, a sub-critical flow can be used to create a high persistent current by reorganizing pre-existing vortex pairs through the Magnus force \cite{ellis_quantum_1993}. In Bose-Einstein condensates, vortices can be created in high numbers by laser stirring \cite{Neely_stirring_2013}.

We show that the proposed device would yield quantized steps in the third-sound mode splitting on the order of 50 Hz, a value within reach of current experimental resolution \cite{harris_laser_2016,sachkou_dynamics_2018}. Note also that since the strength of the vortex-phonon interaction scales inversely with resonator area (see \ref{subsectionanalyticalvortexsoundcoupling}) the magnitude of the splitting can be greatly enhanced by going towards miniature third-sound resonators. This is illustrated in Fig. (\ref{Figure5fig}), which shows the splitting per centered vortex on the (1,2) Bessel mode as a function of resonator radius.   While   only of the order of milli-Hertz for early cm-scale capacitively detected third-sound resonators \cite{ellis_observation_1989, schechter_observation_1998} (red and orange dots), it reaches tens to hundreds of Hz with microtoroidal resonators \cite{harris_laser_2016, sachkou_dynamics_2018} (blue dot), and would attain tens of kHz with micron-radius resonators \cite{gil-santos_scalable_2017,gil-santos_high-frequency_2015,baker_photoelastic_2014} (black dot).

\section{Conclusion}
\label{conclusion}

We have developed finite-element modelling tools to compute the interaction between  any vortex flow and any sound wave, in arbitrary and potentially multiply-connected two-dimensional domains. This capability offers great versatility, applicable to both BEC superfluids and to thin-film superfluid helium. There is a need for numerical techniques to determine vortex and sound velocity fields and their interactions. In both cases, analytical solutions for the vortex flow field only exist if the domain exhibits a high degree of symmetry.  Even if such solutions exist, when departing from simple geometries like a disk, the implementation of the method of images in order to cancel the normal component of the vortex flow on the resonator boundary \cite{lamb_hydrodynamics_1993} becomes challenging, and one needs to rely on conformal mapping techniques \cite{saffman_vortex_1992,gauthier_negative-temperature_2018}. For multiply-connected domains, solutions often require an infinite series of images as the domain possesses two or more boundaries, and analytical solutions are only available for simple limit-cases such as a centered annular domain \cite{fetter_low-lying_1967}.

We verify the validity of our approach by comparing its results to a perturbation theory analysis which we derive in the analytically tractable case of a circular resonator geometry. We derive in this case a useful simple analytical formula,  which can be used to compute the vortex-sound coupling for arbitrary configurations of vortices on a disk, without requiring the vortex flow field.
Understanding precisely how superfluid vortices and persistent currents couple to sound waves -- at the level of a single vortex or circulation quantum -- is  a crucial capability to shed light on the physics of strongly interacting superfluids, and perform continuous non-destructive measurements of vortex dynamics in these systems \cite{sachkou_dynamics_2018}. 

%Understanding strongly interacting superfluids is important for advancing a range of research fields, including quantum computation[], astrophysics[], and the interplay of QFT and gravity[].

The modelling techniques presented here may help shed light on the validity of phenomenological models such as the point-vortex model \cite{simula_emergence_2014,billam_spectral_2015} in superfluids, as well as further our understanding of quantum turbulence \cite{johnstone_order_2018,barenghi_introduction_2014,paoletti_quantum_2011,skrbek_quantum_2011} and energy dissipation in superfluids \cite{ekholm_studies_1980, penanen_model_2002, hoffmann_measurements_2004}.

%A key feature of our work lies in quantifying this interaction in more realistic scenarios - for non-centered vortices and arbitrary, two dimensional resonator shapes. Unlike previous works, it allows the  vortex number and distribution to be extracted independently. Thus, our work provides a theoretical framework for controlling and imaging vortices, and for engineering a dynamical interaction of sound and vortices.

\section*{Acknowledgments}
Authors would like to thank Nick Wyatt for his help in setting up the COMSOL simulation in its early stages,  Prahlad Warszawski, Andrew Doherty, Matt Davis, Rachpon Kalra, Yasmine Sfendla and Andreas Sawadsky for discussions on vortex dynamics, and Gian-Marco Schn\"uriger for discussions on wave dynamics. The authors in particular would like to thank Matt Reeves and David Colas for helpful discussions on vortex dynamics and sound in Bose-Einstein condensates. This work was funded by the U.S. Army Research Office through grant number W911NF-17-1-0310. It was also supported by the Australian Research Council through the Centre of Excellence for Engineered Quantum Systems (EQuS, CE110001013); W.P.B. acknowledges the Australian Research Council Future Fellowship FT140100650. 
C.G.B acknowledges a Fellowship from  the University of Queensland (UQFEL1833877).\\

\appendix
\section{Analytical derivation of the vortex-sound coupling}
\label{sec: analytical expression}

\subsection{Analytical description of the point vortex flow field $\vec{v}_v$}
\label{sectionanalyticalpointvortex}
The streamfunction $\Psi$ of a point vortex of strength $\kappa$ in a 2D plane is $\Psi=-\frac{\kappa}{2\pi}\ln\left(r\right)$. The streamfunction $\Psi$ for the the well-known problem of a point vortex inside a circular domain \cite{lamb_hydrodynamics_1993}, is given in cartesian coordinates by:
\begin{equation}
\Psi = -\frac{\kappa}{2\pi} \left( \ln\left(\sqrt{\left(x-X_1\right)^2+y^2}\right)-\ln\left(\sqrt{\left(x-X_2\right)^2+y^2}\right)\right)
\label{EqPsivortexcirculardomain}
\end{equation}
Here $X_1$ is the radial offset of the vortex (along the $x$ axis), and $X_2=\frac{R^2}{X_1}$ is the radial coordinate of the opposite circulation image-vortex required to enforce no flow accross the resonator boundary \cite{lamb_hydrodynamics_1993}. From the streamfunction $\Psi$, the vortex velocity components are given by:
\begin{equation}
v_{vx}=\frac{\partial \Psi}{\partial y};\qquad \mathrm{and}\qquad
v_{vy}=-\frac{\partial \Psi}{\partial x}
\label{EqvortexvelocityfromPsi}
\end{equation}
Using Eq.(\ref{EqvortexvelocityfromPsi}), we plot in Fig. \ref{Figureanalyticalexplanationfig}(a) the flow streamlines of a vortex offset by R/2 inside a circular resonator of radius R.

\subsection{Analytical description of the sound flow field $\vec{v}_3$}
\label{sectionanalyticalthirdsound}
In the following, for simplicity we derive the analytic expressions for helium thin films (i.e. third-sound waves). However, the analysis can be applied to BECs with a straightforward replacement of variables (see \ref{comparison tables}).

\begin{figure}
\centering
\includegraphics[width=\textwidth]{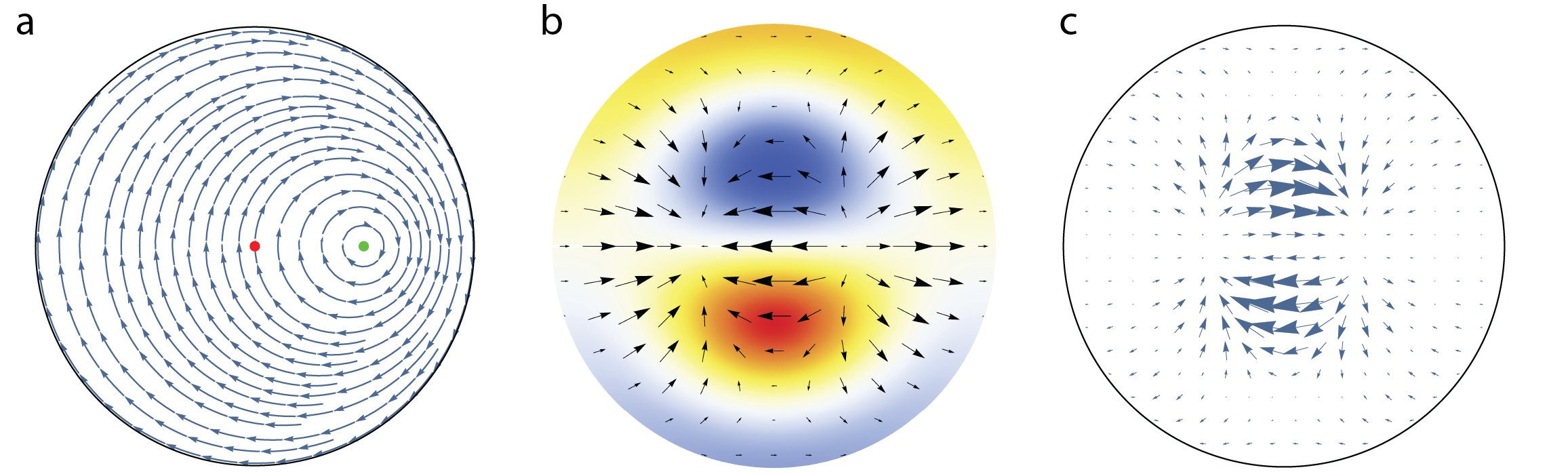}
\caption{(a) Streamlines of $\vec{v}_v\left(\vec{r}\right)$ for a CW vortex (green dot) offset from the origin (red dot) in a circular domain. ($X_1=0.5 R;\, X_2=2R$). (b)  Black arrows represent the instantaneous superfluid flow field $\vec{v}_3\left(\vec{r}\right)$, for a clockwise-rotating ($m=1; \, n=2$) Bessel-mode with free boundary condition. Surface plot shows the associated surface deflection $\eta\left(\vec{r}\right)$ (color code: red = positive, blue=negative). CW nature can be seen by noticing the fluid starting to accumulate ahead of the red peak, where $\vec{\nabla} \cdot \vec{v}_3<0$.
The velocity field is positive under the peaks, negative under the troughs, and irrotational, \textit{i.e.} with 
$\oint \vec{v}_3
%\left(\vec{r}\right)
 \cdot \vec{\rmd l}=0 $ for all contours inside the superfluid. (c) Vector field of $\vec{v}_3\left(\vec{r}\right)\times\eta\left(\vec{r}\right)$. While from symmetry one sees that $\int\hspace{-5pt}\int  \vec{v}_3 \cdot \vec{v}_v=0$, multiplication by the surface deflection profile $\eta\left(\vec{r}\right)$ leads to a non-zero energy shift of  the CW/CCW third-sound waves, see Eq.(\ref{EqDeltaE4}).}
\label{Figureanalyticalexplanationfig}
\end{figure}

The complex surface displacement amplitude $\eta$ of a travelling superfluid third-sound wave (or alternatively the sound-induced density fluctuations for a BEC) in a circular domain are given by \cite{baker_theoretical_2016}:
\begin{equation}
\eta\left(r,\theta, t\right)=\eta_0\,J_m\left(\zeta_{m,n} \frac{r}{R}\right)e^{i\left(m\theta \pm \Omega\,t\right)}.
\label{Eqetathirdsound}
\end{equation}
The associated superfluid flow speed $\vec{v}_3$ is given by:
\begin{equation}
\vec{v}_3=\pm \frac{i \,c_3^2}{\Omega\, h_0}\vec{\nabla}\eta %\left(\vec{r}\right) 
\label{Eqv3}
\end{equation}
with `+' and `-' signs respectively corresponding to the CW and CCW travelling cases, as in Eq.(\ref{Eqetathirdsound}).
Note that while the motion of a solid circular membrane would also be given by Eq.(\ref{Eqetathirdsound}), its velocity would be different to Eq.(\ref{Eqv3}), leading to dramatically different effective mass scalings \cite{baker_theoretical_2016}. The surface displacement profile $\Re \left(\eta\right)$ and instantaneous velocity field $\Re \left( \vec{v}_3 \right)$ are plotted in Fig. \ref{Figureanalyticalexplanationfig}(b), for a CW $(m=1, \,n=2)$ Bessel mode with free boundary conditions.
While such a third-sound mode flow is irrotational and therefore not associated with any circulation ($\oint\vec{v}\cdot\vec{\rmd l}=0$ for any closed loop inside the superfluid), it is associated with a net mass flow (in the CW case, there is more fluid moving clockwise under the wave peak (red) than counter-clockwise under the trough (blue), and similarly there is net CCW fluid motion for the CCW mode). It is this net mass flow which couples to the vortex field, and results in a higher kinetic energy for the sound mode travelling in the same direction as the vortex flow. This argument is developed in the analytical splitting calculation detailed below.

\subsection{Analytical description of vortex-sound coupling}
\label{subsectionanalyticalvortexsoundcoupling}
%This perturbation theory approach is derived in the quasi-static regime where the motion of the vortices during a sound oscillation period is negligible. This is valid in the limit
%of pinned vortices \cite{ellis_quantum_1993, ellis_excitation_1998, hoffmann_measurements_2004} or low vortex densities where the velocity of the flow field due to neighboring and image vortices is significantly less than the speed of %sound\footnote{The orbit period T for a single vortex offset from the center of a circular resonator of radius R by a distance x is $T=\frac{4\pi^2}{\kappa}\left( R^2-x^2 \right)$. For devices with $R\simeq 10^{-5}$ m, this corresponds to typical Hz %orbit frequencies compared to typical $10^5$ Hz third-sound frequencies \cite{harris_laser_2016}.}.

Here we derive an analytical expression for the frequency splitting experienced by a third-sound mode due to a vortex inside a circular resonator, and show good agreement with the results of the FEM simulations shown in Figure \ref{Figure2fig}. This is a valid approximation to the FEM-model if the change in mode shape due to vortices is small.

\begin{figure}
\centering
\includegraphics[width=\textwidth]{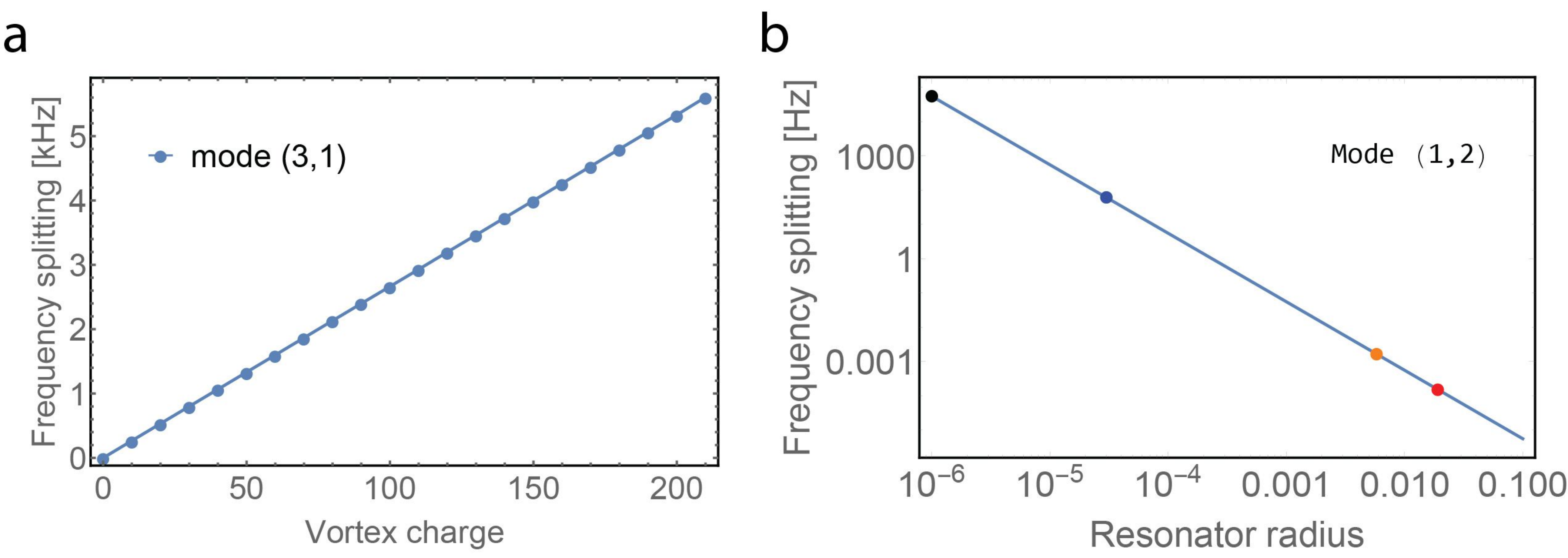}
\caption{(a) FEM simulation of frequency splitting of the $(m=3,\,n=1)$ Bessel mode due to a centered vortex, whose charge is increased from $\kappa$ to  $>200\, \kappa$, displaying linearity over that range. (b) Splitting per centered vortex for the $(m=1,\,n=2)$ Bessel mode with free boundary conditions, as a function of resonator radius. Experimental devices shown in red\cite{schechter_observation_1998} and orange \cite{ellis_observation_1989} correspond to cm-scale capacitively detected third-sound waves. Blue dot corresponds to an optical WGM  microtoroid resonator\cite{harris_laser_2016, sachkou_dynamics_2018}. Black dot shows  two additional orders of magnitude improvement over current state-of-the-art can be achieved by going to micron-radius WGM resonators \cite{gil-santos_scalable_2017,gil-santos_high-frequency_2015}. }
\label{Figure5fig}
\end{figure}

The kinetic energy difference $\Delta E\left(t\right)$ between a sound wave moving with- or against the flow of a quantized vortex is given by:
%In the presence of a vortex, the kinetic energy difference between co- and counter-rotating third sound sound modes
\begin{equation}
\hspace{-80pt}\Delta E\left(t\right)=\frac{1}{2} \, \rho\int_{\theta=0}^{2\pi}\int_{r=0}^R\int_{z=0}^{h_0+\eta\left(r,\,\theta,\,t\right)} \left(\left|\left| \vec{v}_3\left(\vec{r}, \, t \right)+\vec{v}_v\left(\vec{r}\right)\right|\right|^2 -\left|\left| \vec{v}_3\left(\vec{r}, \, t\right)-\vec{v}_v\left(\vec{r}\right)\right|\right|^2 \right)r\, \rmd r \,\rmd\theta \, \rmd z
\label{EqdeltaEnumber1}
\end{equation}
This general expression works for any sound mode and any vortex position.
Making the reasonable assumption that $\vec{v}_3$ and $\vec{v}_v$ are independent of z, as the inviscid nature of the superfluid precludes any in-plane vorticity and does not require cancellation of the horizontal velocity at z=0 ({\it no-slip} boundary), Eq.(\ref{EqdeltaEnumber1}) becomes:
%\begin{equation}
%\hspace{-40pt}\Delta E=2\rho\int_{\theta=0}^{2\pi}\int_{r=0}^R   \vec{v}_3\left(\vec{r}\right)\cdot \vec{v}_v\left(\vec{r}\right) \left(h_0+ \eta\left(r,\theta\right)\right) r \, \rmd r \, \rmd\theta 
%\end{equation}
\begin{equation}
\hspace{-40pt}\Delta E\left(t\right)=2\rho\int_{\theta=0}^{2\pi}\int_{r=0}^R   \vec{v}_3\left(r,\theta, \, t\right)\cdot \vec{v}_v\left(r,\theta\right) \left(h_0+ \eta\left(r, \theta, t \right)\right) r \, \rmd r  \, \rmd\theta 
\label{EqdeltaEnumber2}
\end{equation}
Since both $v_{3x}$ and $v_{vx}$ as well $v_{3y}$ and $v_{vy}$ are functions of $\theta$ of different parity (see Figure \ref{Figureanalyticalexplanationfig}), $\int\hspace{-5pt}\int  \vec{v}_3 \cdot \vec{v}_v=0$, and equation (\ref{EqdeltaEnumber2}) becomes:  
\begin{equation}
\hspace{-40pt}\Delta E\left(t\right)=2\rho\int_{\theta=0}^{2\pi}\int_{r=0}^R   \vec{v}_3\left(r,\theta, t\right)\cdot \vec{v}_v\left(r,\theta\right) \eta\left(r,\theta, t\right) r \, \rmd r \, \rmd\theta 
\label{EqDeltaE4}
\end{equation}
This is essentially a form of surface-averaged Doppler shift, weighted by the displacement amplitude $\eta$ of the mode. Next, we consider the time-averaged energy difference $\left<\Delta E\right>$, averaged over a sound oscillation period T:
\begin{equation}
\hspace{-40pt}\left<\Delta E\right>= \frac{1}{T} \int_0^T \Delta E\left( t \right) \rmd t= 2\,\rho \int_r \int_{\theta} r \, \rmd r \, \rmd\theta \left(v_{v\,r} \, \frac{1}{T} \int_0^T v_{3\,r} \, \eta \, \mathrm{d} t \right) + \left(v_{v\,\theta} \, \frac{1}{T} \int_0^T v_{3\,\theta} \, \eta \, \mathrm{d} t \right)  
\label{EqDeltaE5}
\end{equation}
where we have broken down $\vec{v}_3$ and $\vec{v}_v$ into their radial and angular components, respectively $v_{3\,r}$ and $v_{3\,\theta}$, and $v_{v\,r}$ and $v_{v\,\theta}$.
From Eqs.(\ref{Eqetathirdsound})\&(\ref{Eqv3}), we note that $v_{3\,r}$ and $\eta$ are out-of-phase, while $v_{3\,\theta}$ and $\eta$ are in phase. The first integral over time in Eq.(\ref{EqDeltaE5}) reduces therefore to zero, while the second integrates to $\frac{1}{2} \left| v_{3\,\theta} \right| \left| \eta \right|$. We therefore get from Eq.(\ref{Eqetathirdsound}) and Eq.(\ref{Eqv3}):

\begin{equation}
\left<\Delta E\right>=\frac{\rho \, m\,c_3^2}{\Omega\,h_0} \int_{r=0}^R  r\, \rmd r \,  \eta_0^2\, \frac{J_m^2\left(\zeta_{m,n} \frac{r}{R}\right)}{r} \int_{\theta=0}^{2\pi} \,v_{v \, \theta} \, \rmd \theta
\end{equation}
which we rewrite, with $\eta\left(r\right)=\eta_0 \,J_m\left( \zeta_{\mathrm{m,n}} \, \frac{r}{R}\right)$, as:

\begin{equation}
\left<\Delta E\right>=\frac{\rho \, m\,c_3^2}{\Omega\,h_0} \int_{r=0}^R   \frac{\rmd r}{r}   \,  \eta^2\left(r\right) \int_{\theta=0}^{2\pi} \,v_{v \, \theta} \, r\,\rmd \theta
\label{EqDeltaE7}
\end{equation}
We notice here that the integral over $\theta$ corresponds to a closed contour integral $\oint \,\vec{v}_{v} \cdot \rmd \vec{l}$, where the contour is a circle of radius $r$ centered at the origin. From Eq.(\ref{eqvortex}), we know that the value of this contour integral is zero if it does not enclose the vortex core, and $\kappa$ if it does. The transition occurs for $r=\mathrm{offset}$, the radial offset of the point vortex. We can therefore rewrite Eq.(\ref{EqDeltaE7}) with a modified radial integration lower bound:

\begin{equation}
\left<\Delta E\right>=\frac{\rho \, m \, c_3^2 \,\kappa}{\Omega\, h_0} \int_{r=\mathrm{offset}}^R  \frac{\rmd r}{r}   \,  \eta^2\left(r\right)
\end{equation}
Since for a harmonic oscillator $E$ is proportional to $\Omega^2$, $\frac{\Delta E}{E}=2\frac{\Delta\Omega}{\Omega}$ and the splitting  $\Delta f$ (in Hz) equals:
\begin{equation}
\Delta f= \frac{\Omega}{4\pi} \frac{\Delta E}{E},
\end{equation}
%\begin{equation}
%E=\frac{1}{2}\int \rho \,v^2\left(\vec{r}\right) \rmd^3\left(\vec{r}\right)=\frac{3 \pi \, \rho \, \alpha_{\mathrm{vdw}}}{2\,h_0^4} \int_{0}^R  \eta^2\left(r\right) r\, \rmd r
%\end{equation}
%for $m>0$ \cite{baker_theoretical_2016}
with the kinetic energy E of the third-sound mode, for $m>0$, given by \cite{baker_theoretical_2016}: 
\begin{equation}
E=\frac{1}{2}\int \rho \,v^2\left(\vec{r}\right) \rmd^3\left(\vec{r}\right)=\frac{\pi \, \rho \, c_3^2}{2\,h_0} \int_{0}^R  \eta^2\left(r\right) r\, \rmd r
\label{Eqenergythirdsound}
\end{equation}
Combining Eqs.(\ref{EqDeltaE7}) and (\ref{Eqenergythirdsound}), we recover the result shown in Eq.(\ref{eqsplittingmaintext}) of the main text:
\begin{equation}
\Delta f=\frac{\kappa\,m}{2\, \pi^2}\, \, \frac{\int_{\mathrm{offset}}^{R}\frac{\rmd r}{r}\,\eta^2\left(r\right)}{\int_{\mathrm{0}}^{R} \rmd r \,r\,\eta^2\left(r\right)} 
\label{Eqsplittingfinalsupplements}
 \end{equation}
We note that, as expected, the splitting does not depend on the superfluid parameters (film thickness, density), and that it is linear in vortex flow field (see Eq.\ref{EqdeltaEnumber2}), such that the splitting obeys the superposition principle, whereby the splitting due to an ensemble of vortices is equal to the sum of the splittings per vortex calculated individually. The result Eq.(\ref{Eqsplittingfinalsupplements}) holds for both superfluid helium thin films and Bose-Einstein condensates, with $\eta$ being the film thickness perturbation/density perturbation, respectively. We numerically verify this result in the FEM simulations, where linearity is generally maintained up to large vortex charges on the order of $\sim 10^{2}\,\kappa$, as shown in Figure \ref{Figure5fig}(a) and (b). Note that Eq.(\ref{Eqsplittingfinalsupplements}) does not diverge as the vortex offset tends to $0$, as $\eta\left(0\right)=0$ for all $m>0$ Bessel modes. Interestingly, due to the contour-integral identity used in Eq.(\ref{EqDeltaE7}), the final result does not require any knowledge of the vortex flow field $\vec{v}_v$.

\begin{figure}
\centering
\includegraphics[width=.8 \textwidth]{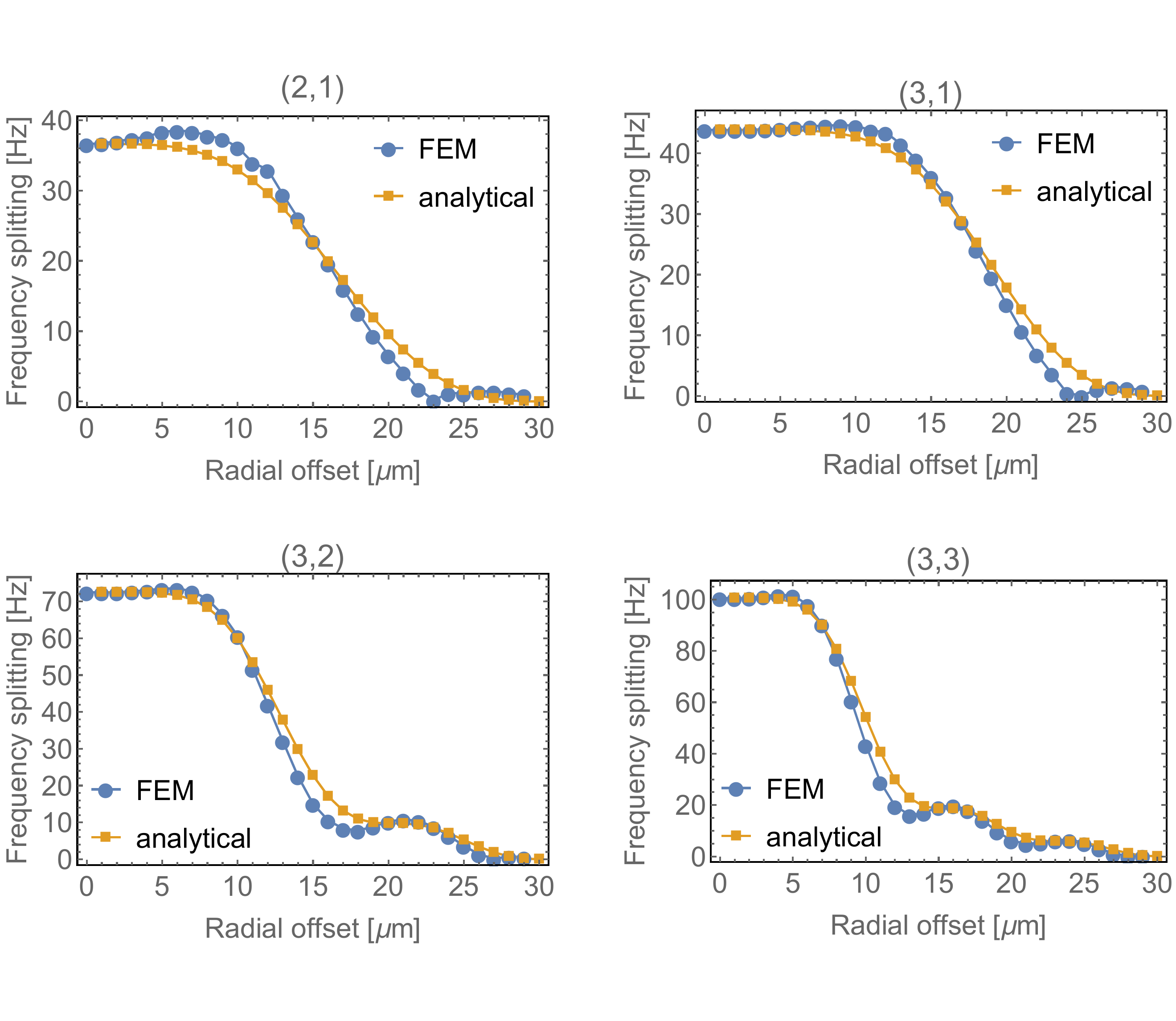}
\caption{Comparison between the results of the FEM simulations and the analytical approach (Eq.(\ref{Eqsplittingfinalsupplements})) for four different Bessel modes labelled by their $(m,n)$ order, showing good agreement between both methods without any scaling parameter.
(Resonator dimension $R=30$ microns, fixed boundary conditions). 
Some small quantitative differences between both solutions remain. For instance from Eq. \ref{Eqsplittingfinalsupplements}, the analytical splitting has to be a monotonically decreasing function of the radial offset, while the FEM calculation shows some regions of increased splitting with radial offset. We ascribe these differences to vortex-induced changes in the eigenmode shape (see Figure \ref{Figure1fig}(c)), which are not taken into account in the perturbative analytical approach.

}.
\label{Figure7fig}
\end{figure}

\section{Derivation of linearized equations for an ideal gas}
\label{seclinac}
 For the ideal gas, mass conservation and momentum conservation read, respectively\cite{rienstra_introduction_2004}:

\begin{equation}
\frac{d\rho}{dt}=-\vec{\nabla} \cdot (\rho\,\vec{u})
\label{accontinuity}
\end{equation}

and

\begin{equation}
\frac{d\vec{u}}{dt}+(\vec{u}\cdot\vec{\nabla})\vec{u}=-  \frac{1}{\rho}\vec{\nabla} p,
\label{acEuler}
\end{equation}
where $\vec{u}(\vec{r},t)$ is the flow velocity, $\rho$ is the gas density and $p$ the gas pressure.  Isentropic flow (i.e. the gas is in thermal equilibrium at all times) for an ideal gas implies\cite{rienstra_introduction_2004}:

\begin{equation}
p=\gamma RT\rho \quad {\rm and} \quad c^2=\gamma R\,T,
\label{isentropic}
\end{equation}
where $R$ is the specific gas constant, $T$ is the gas temperature and $c$ is the speed of sound. We insert Eq. (\ref{isentropic}) in Eq. (\ref{acEuler}) and  linearize for small density fluctuations, $\rho(\vec{r})=\rho_0+\alpha(\vec{r})$ with $\alpha \ll \rho_0$, and recover Eqs. (\ref{tslincontinuity}) and (\ref{tslinEuler}) in the main text.
\section{Comparison tables}
\label{comparison tables}
Here we show how quantities and equations from 2D-electrostatics can be mapped to vortex-induced flow fields, and how acoustics of an ideal gas is mapped to third-sound dynamics.

\begin{table}[h!]
\center
\begin{tabular}{c | c} 

2D-electrostatics & vortices \\

\hhline{=|=}
electric displacement field & velocity field \\
$\vec{D}(\vec{r})$[C/m$^2$]  & $\vec{v}_v(\vec{r})$[m/s]\\
\hline
electric line charge & circulation quantum \\
 $Q$[C/m] & $ \kappa$[m$^2$/s]\\
\hline
 Gauss's law & vortex flow equation\\
$\oint \vec{D}\cdot d\vec{n}=Q$ & $\oint \vec{v}_v \cdot d\vec{l} = \kappa$\\
\hline
perfect electric conductor ({\it ground}) & tangential flow boundary \\
$\vec{D} \times \vec{n} = 0$ & $\vec{v}_v \cdot \vec{n} = 0$\\
\hline
\end{tabular}
\caption{electrostatics and vortex flow field. The system is invariant under $z$-translation, hence we use units and equations in two dimensions.}
\label{tab:electrostatics table}
%\vspace{2cm}
\end{table}

\begin{table}[h!]
%\center
\hspace{-1cm}
\begin{tabular}{c | c | c}
 
2D-acoustics & sound in 2D-BEC  & third-sound dynamics \\

\hhline{=|=|=}
density perturbation &density perturbation &third-sound amplitude \\
$\alpha(\vec{r},t)$[kg/m$^2$]  &$\eta(\vec{r},t)$[kg/m$^2$] & $\eta(\vec{r},t)$[m]\\
\hline
static density &static density &unperturbed film height \\
 $\rho_0$[kg/m$^2$] & $\rho_0$[kg/m$^2$]  & $h_0$[m]\\
\hline
background flow  & irrotational vortex flow  & irrotational vortex flow  \\
 $u_0(\vec{r})$[m/s] &$v_v(\vec{r})$[m/s] & $v_v(\vec{r})$[m/s]\\
\hline
irrotational flow velocity &sound flow velocity &third-sound flow velocity \\
 $\delta\vec{u}(\vec{r},t)$[m/s] &$\vec{v}_s(\vec{r},t)$[m/s]& $\vec{v}_3(\vec{r},t)$[m/s]\\
\hline
static pressure &atom-atom coupling & linearized VdW coefficent\\
$p_0$[J/m$^2$] &$g_{\rm BEC}$[Jm$^2$] & $g=\frac{3\alpha_{\mathrm{vdw}}}{h_0^4}$[m/s$^2$]\\
\hline
speed of sound (acoustics) & Bogoliubov sound velocity &speed of sound (thin film)\\
$c=\sqrt{\gamma RT}$ [m/s] & $c=\sqrt{g_{\rm BEC}\cdot \rho_0/M^2}$[m/s]& $c_3=\sqrt{g\cdot h_0}$ [m/s]\\
\hline
fixed wall boundary &fixed wall boundary& free boundary \\
$\vec{u} \cdot \vec{n} = 0 $ & $\vec{v} \cdot \vec{n} = 0$&$\vec{v} \cdot \vec{n} = 0$\\
\hline
fixed pressure boundary &fixed density boundary  &fixed boundary \\
$p=p_0$ & $\eta=0$& $\eta=0$\\
\hline
continuity equation (acoustics) &continuity equation (BEC) &continuity equation (thin film)\\
$\dot{\alpha}=-\rho_0 \vec{\nabla} \cdot \vec{u} - \vec{u}\vec{\nabla}\alpha$ &$\dot{\eta}=-\rho_0\vec{\nabla} \cdot \vec{v} - \vec{v}\vec{\nabla}\eta$&$\dot{\eta}=-h_0\vec{\nabla} \cdot \vec{v} - \vec{v}\vec{\nabla}\eta$\\
\hline
linearized Euler (acoustics) & linearized Euler (BEC)  &linearized Euler (thin film)\\
$\dot{\vec{u}}+(\vec{u}\cdot\vec{\nabla})\vec{u} = -\frac{\gamma RT}{\rho_0}\vec{\nabla} \alpha$& $\dot{\vec{v}}+(\vec{v}\cdot\vec{\nabla})\vec{v}=- \frac{1}{M} \vec{\nabla}\big(U+\frac{\eta}{M} g_{\rm BEC}\big)$ & $\dot{\vec{v}}+(\vec{v}\cdot\vec{\nabla})\vec{v}=- g \vec{\nabla} \eta$\\
\end{tabular}
\caption{acoustics, sound dynamics in a Bose-Einstein condensate in the Thomas-Fermi limit at zero temperature, and third-sound dynamics on a helium thin film. As in table \ref{tab:electrostatics table}, a two-dimensional system is described.}
\label{tab:acoustics table}
\end{table}

\section{Supplementary information}

\subsection{Boundary conditions}
\label{sec:boundary conditions}
 In order to solve differential equations on the surface of a two-dimensional resonator, constraints at the boundary have to be specified.  Depending on the type of confinement provided by the resonator, the boundary for third sound can be described either by a fixed (\textit{`Dirichlet'}) or a free (\textit{`Neumann'}) boundary condition. A fixed boundary condition $\eta= 0$ allows flow in and out of the resonator and the film height at the boundary is fixed to the equilibrium film height. 
%This case describes a resonator with an edge that is continually coated with a superfluid film, such as realized e.g. in ref.\cite{harris_laser_2016}. 
The free boundary condition $\vec{v}_3\cdot \vec{n} = 0$, where $\vec{n}$ is the normal vector on the boundary, allows film height fluctuations at the boundary and prohibits flow in or out of the resonator. In COMSOL, for an ideal gas, the free boundary condition corresponds to a \textit{rigid wall}, where volume is conserved and the gas pressure can oscillate freely at the boundary. The fixed boundary condition corresponds to \textit{fixed pressure}, where the gas pressure is fixed at the boundary and the gas can freely flow in and out of the domain\cite{noauthor_acoustics_nodate}.  The vortex flow is tangential to the boundary, $\vec{v}_v \cdot \vec{n} = 0$. In the electrostatics analogue, this translates to an electric field which is exactly perpendicular to the boundary, with no tangential component. This corresponds to a perfect electric conductor at the boundary and can be realized by choosing the \textit{ground} - boundary condition in COMSOL\cite{noauthor_ac/dc_nodate}. In order to model a quantized circulation $n \times \kappa$ around a topological defect in the structure, the \textit{floating potential} boundary condition with built in charge $Q=n\times\kappa$ must be chosen.  This boundary condition enforces an electric field  orthogonal to the boundary everywhere (due to the equal potential on the boundary), as well as the condition:
 
\begin{equation}
\oint \vec{D}\cdot \vec{n}\, dl =Q.
\end{equation}
Upon the substitution of Eq.(\ref{EqpermutationDgoestov}), this corresponds to a superfluid flow always parallel to the topological defect boundary (\textit{i.e.} no fluid inflow or outflow), and the quantized circulation condition:

\begin{equation}
\oint \vec{v}\cdot \rmd\vec{l} =n\, \kappa
\end{equation}

\subsection{Notes on implementation in COMSOL\textsuperscript{\textregistered} multiphysics}

In the following we describe how superfluid helium thin film can be modelled using the FEM solver COMSOL\textsuperscript{\textregistered} multiphysics 5.0.

A 2D model is set up. The {\it Electrostatics(es) module} is used to simulate vortices and a {\it stationary} study is created. The resonator outer boundary is set to {\it ground}. The circulation quantum $\kappa$ is defined with adjusted SI-units ($\frac{\rm m}{\rm s^2}\rightarrow \frac{\rm C}{\rm m}$). At each position where a clockwise vortex is to be modelled, a {\it line charge (out-of-plane)} of $Q_{\rm L}=\kappa$ is inserted. A counter-clockwise vortex can be modelled by replacing $\kappa\rightarrow -\kappa$.

To model third sound, the {\it Aeroacoustics $\rightarrow$ Linearized Euler, Frequency Domain(lef) module} is added and a {\it Eigenfrequency} step is included in the study. In the first, stationary, study step, only the electrostatics interface is solved for, whereas in the second step the {\it Eigenfrequency} solver is applied to the acoustics interface. Parameters $\rho_{\rm sf}$, $A_{\rm vdw}$, $h_0$ and $g=3\alpha_{\rm vdw}h_0^{-4}$\cite{tilley_superfluidity_1990} are defined to set the superfluid density ($145 \frac{\rm kg}{\rm m^3}$ for superfluid helium\cite{donnelly_observed_1998}), the Hamacker constant of the substrate ($2.6\cdot 10^{-24}\frac{\rm m^5}{\rm s^2}$ for silica\cite{baker_theoretical_2016}), the film thickness and the linearized Van-der-Waals acceleration, respectively. The product $RT\gamma=c^2$ is set to $g\cdot h_0$ for a uniform superfluid film. Alternatively, a spatially varying function can be defined to reflect a non-uniform film thickness.  The boundary condition is set to either \textit{rigid wall} or \textit{fixed pressure} (see section \ref{sec:boundary conditions}). 

In order to include vortices defined in the {\it Electrostatics(es) module}, a critical velocity is defined ($v_{\rm crit}\approx 60$ $\rm m/s$ for superfluid helium\cite{landau_theory_1941}), and the acoustic background flow field $\vec{u}_0$ is set to:
\begin{equation}
\left(
\begin{array}{c}
 u_{0,x} \\
 u_{0,y} \\
\end{array}
\right)=\left(
\begin{array}{c}
 D_y \\
 -D_x \\
\end{array}
\right),
\end{equation}
(where D is the electric displacement field solved for in the first step stationary solver), and truncated at $u_{0,max} =v_{\rm crit}$. This vortex background flow field is treated as constant in time. The FEM-solver computes the velocity perturbation $\delta\vec{u}(t)$, corresponding to the third sound mode, to the total flow: $\vec{u}(\vec{r},t)=\vec{u}_0(\vec{r})+\delta\vec{u}(\vec{r},t)$. 

The gas density perturbation $\alpha(\vec{r},t)$ calculated in COMSOL\textsuperscript{\textregistered} can be converted to third sound amplitude by a normalization factor: $N=\alpha(\vec{r},t)/\eta(\vec{r},t)$. It is extracted from

\begin{equation}
E_{\rm pot,3}(\eta)=k_{\rm B}T_{\rm mode},
\end{equation}
where $T_{\rm mode}$ is the mode temperature. An analytical expression for the potential energy of a third sound mode $E_{\rm pot,3}$ is given in ref.\cite{baker_theoretical_2016}. For the simple case of a uniform film thickness and free boundary conditions the conversion is given by

\begin{equation}
N=\frac{3A_{\rm vdw}\rho_{\rm sf}}{2d^4k_{\rm B}T_{\rm mode}} \cdot \int d^2\vec{r}\, \rho(\vec{r}).
\end{equation}

$T_{\rm mode}$ is the effective temperature of the sound mode, which when thermalized with its environment corresponds to the fridge temperature. It can also be tuned through optomechanical laser heating/cooling \cite{harris_laser_2016}, increased through laser absorption heating \cite{mcauslan_microphotonic_2016} or electrical excitation \cite{ellis_quantum_1993}.

\section*{References}

\bibliographystyle{unsrt} %  iopart-num

\bibliography{referencessuperfluidvortices_no_duplicates_v5}

\end{document}